\documentclass[9pt,conference]{hahaha}
\IEEEoverridecommandlockouts
\usepackage[noadjust]{cite}

\usepackage{listings}
\usepackage{booktabs} 
\usepackage{color}
\usepackage{multirow}
\usepackage{amsthm}
\usepackage{array}
\usepackage[pdftex]{graphicx}
\usepackage{comment}
\graphicspath{{./fig/}}
\usepackage{color}
\usepackage[bookmarks=false]{hyperref}
\usepackage{xcolor}
\usepackage{subfigure}
\usepackage{cite}
\usepackage{amsmath,amssymb,amsfonts}
\usepackage{algorithmic}
\usepackage{graphicx}
\usepackage{textcomp}
\usepackage{xcolor}
\usepackage[linesnumbered,ruled,vlined]{algorithm2e}

\SetCommentSty{mycommfont}
\usepackage{mathtools}

\usepackage{subfigure}
\def\BibTeX{{\rm B\kern-.05em{\sc i\kern-.025em b}\kern-.08em
		T\kern-.1667em\lower.7ex\hbox{E}\kern-.125emX}}

\begin{document}
	
	\title{AutoCellLibX: Automated Standard Cell Library Extension Based on Pattern Mining}	
	
	\author{\IEEEauthorblockN{Tingyuan Liang\IEEEauthorrefmark{1}, Jingsong Chen\IEEEauthorrefmark{2}, Lei Li\IEEEauthorrefmark{2} and Wei Zhang\IEEEauthorrefmark{1}}
		\IEEEauthorblockA{\textit{\IEEEauthorrefmark{1}ECE Department, Hong Kong University of Science and Technology;} \textit{\IEEEauthorrefmark{2}Huawei Technologies Co., Ltd. Shenzhen, China} \\
			tliang@connect.ust.hk, chenjingsong5@huawei.com, lilei291@huawei.com, eeweiz@ust.hk}
	}
	
	\maketitle	
	\begin{abstract}
		Custom standard cell libraries can improve the final quality of the corresponding VLSI designs but properly customizing standard cell libraries remains challenging due to the complex characteristics of the VLSI designs. This paper presents an  automatic standard-cell library extension framework, AutoCellLibX. It can find a set of standard cell cluster pattern candidates from the post-technology mapping gate-level netlist, with the consideration of standard cell characteristics and technology mapping constraints, based on our high-efficiency frequent subgraph mining algorithm. Meanwhile, to maximize the area benefit of standard cell customization for the given gate-level netlist, AutoCellLibX includes our proposed pattern combination algorithm which can iteratively find a set of gate-level patterns from numerous candidates as the extension part of the given initial standard cell library. To the best of our knowledge, AutoCellLibX is the first automated standard cell extension framework that closes the optimization loop between the analysis of gate-level netlist and standard cell library customization for VLSI design productivity. The experiments with FreePDK45 library and benchmarks from various domains show that AutoCellLibX can generate the library extension with up to 5 custom standard cells within 1.1 hours for each of the 31 benchmark designs and the resultant extension of the standard cell library can save design area by 4.49\% averagely.
	\end{abstract}
	
	\begin{IEEEkeywords}
		standard cell synthesis, dynamic standard cell library, design technology co-optimization, frequent sub-graph mining
	\end{IEEEkeywords}
	
	\section{Introduction}
	
	Digital VLSI designs are getting much larger and more complex as manufacturing technologies are making rapid progress. Standard cell-based designs exploit libraries of small predefined building blocks called standard cells, which facilitate the productivity and reliability of the VLSI design flow. Accordingly, standard cell libraries have a fundamental impact on the quality of VLSI designs, e.g., power, performance, area, and cost (PPAC)~\cite{PD}. 
	
	\subsection{Background of Standard Cell Merging}
\begin{figure}[t]
		\centering
		\includegraphics[width=\linewidth]{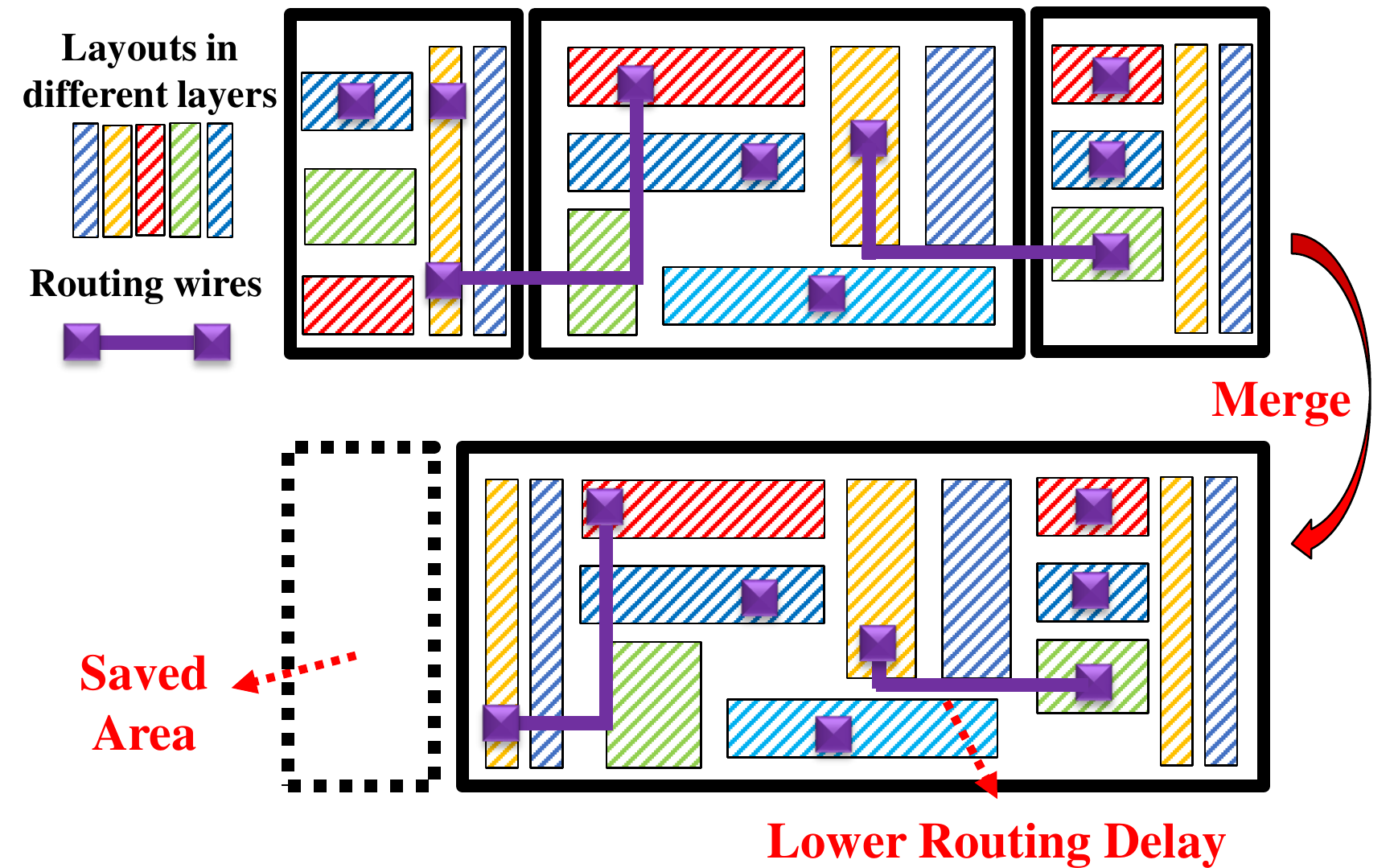}
		\caption{Simplified example of standard cell merging: it shows the advantages of standard cell merging which enables transistor-level optimization.} 
		\label{merging}
	\end{figure}

	Since commonly-used standard cell libraries cannot meet all the requirements in some special scenarios~\cite{domainASIC,domainFPGA,ALPINE,pattern}, as an alternative solution, academia~\cite{CELLERITY,cellTK,ASTRAN,Libra,ASAP7,gateMerging,iTPlace,li2020mcell,DTCO,SPR,NCTUcell,7nm2019,Bonncell,PROBE2} and industry~\cite{postlayoutPattern,specificComplex,pattern,IBM,NVCell} take continuous efforts to extend standard cell libraries with custom standard cells for their technology nodes and domain-specific designs. One of the potential sources of these custom standard cells is \textbf{\emph{standard cell merging}}, which merges several existing standard cells into a new one with an optimized layout, as a simplified example shown in Fig.~\ref{merging}. It can benefit the design flow from two aspects:
	\begin{itemize}
		\item It enlarges the back-end solution space of the VLSI design, e.g., placement and routing,  and hence enables some design goals which cannot be achieved based on the original standard cell library, via transistor-level optimizations like diffusion sharing and in-cell signal routing~\cite{ASTRAN}.
		\item It shrinks the problem scale of the VLSI back-end tools by reducing the number of gates in the post-technology mapping gate-level netlist and enforcing pre-defined relative position constraints. These factors facilitate the tools to converge at better results~\cite{jianliPlacer}.
	\end{itemize}
	
	Standard cell merging provides noticeable optimization potentials but it is critically challenging since numerous factors, e.g., the design context, transistor layout, design rules and expected PPAC metrics, should be considered to realize a beneficial library. First, as for the design context, designers should identify the characteristics of the target design to locate the optimization opportunities and design pattern mining is one of the promising approaches. Second, the transistor network and layout should be designed under the constraints of design rules and PPAC metrics, which is usually called standard cell layout synthesis. In this paper, we propose a fully automatic standard-cell library extension framework, AutoCellLibX which can analyze characteristics of the target gate-level netlist and extent an initial standard cell library with custom complex standard cells to minimize the area cost. Related works and challenges are discussed in Section~\ref{background} and ~\ref{challenges} respectively.

	\subsection{Related Works}\label{background}
	
	\subsubsection{Standard Cell Layout Synthesis}
	This workflow takes in three types of inputs~\cite{CELLERITY,PROBE2}: (A) cell architecture settings, e.g. the number of routing tracks, and the location of power rails; (B) design rules, e.g., width/spacing rules for all mask layers depending on cell architecture; and (C) transistor netlist indicating the interconnection of sized library cells. With these inputs, standard cell layout synthesis will conduct transistor placement, in-cell routing, and design rule checking. The general objectives of this synthesis flow are to improve the performance and lower the consumption of area/power. 
	
	Cellerity~\cite{CELLERITY} was proposed in 1997 as a comprehensive standard cell layout synthesis flow. It consists of simulated annealing transistor placement, Echelon in-cell router~\cite{Echelon}, and SQUEEZE layout compactor~\cite{SQUEEZE}. In 2014, ASTRAN~\cite{ASTRAN}, an open-source standard cell layout synthesis framework, was released and it realized promising results based on a cell layout compaction methodology using MILP. Moreover, some solutions~\cite{ye2015standard,xu2016parr,yu2020pin} related to layout regularity and pin access optimization have been done. Recently, a series of researches targeting sub-10nm technology nodes~\cite{SPR,NCTUcell,Bonncell,7nm2019,PROBE2} were presented, most of which utilized SMT solvers to optimize the layout of cell designs and,  ~\cite{SimultaneousLayout} realized efficient simultaneous transistor folding and placement. These frameworks focus on the transistor layouts, and the analysis of the overall gate-level netlist and the input of customization are left to the manual effort.
	
	\subsubsection{Design Pattern Mining}
\begin{figure}[t]
		\centering
		\includegraphics[width=0.9\linewidth]{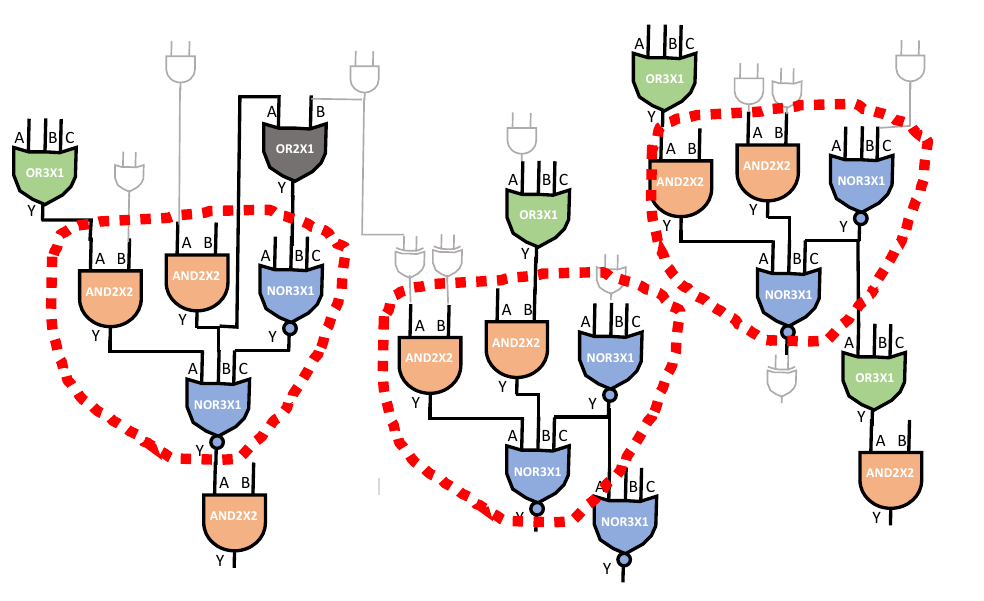}
		\caption{Example of Frequent Subgraph Mining: It can find the frequently-recurring local logic structures (i.e., design patterns) in a large gate-level design for standard cell customization} 
		\label{FSM_intro}
	\end{figure}

	\textbf{\emph{Isomorphic design patterns}} are defined as the identical or functionally equivalent logic structures which recur frequently in the entire design, as an example shown in Fig.~\ref{FSM_intro}. Most styles of the manual design flow, design architecture, synthesis, or mapping have a bias to generate recurring regular patterns in the gate-level netlists~\cite{pattern}. Such regularity can guide the standard cell design since frequently-recurring patterns imply the high coverage and reuse of the corresponding custom standard cell designs.
	
	Accordingly, design pattern mining is to identify which part of the gate-level netlist should be replaced by custom standard cell designs. It's hard since the netlists are large and complex and related solutions were not provided in the aforementioned previous works~\cite{CELLERITY,cellTK,ASTRAN,Libra,ASAP7,gateMerging,iTPlace,DTCO,SPR,NCTUcell,7nm2019,Bonncell,PROBE2,postlayoutPattern,specificComplex,pattern,IBM,NVCell}. In some works on synthesis, placement, and routing, researchers have tried to extract template-based regularity from gate-level netlists, e.g., datapath~\cite{datapath1,datapath2,datapath3,datapath4} or array~\cite{array1}, but it is hard for these template-based solutions to handle the general gate-level netlists with complex interconnections. Moreover, they did not check the isomorphism between the patterns.
	
	Frequent subgraph (pattern) mining (FSM) is a hot topic of computer science with some matured solutions like ~\cite{apriori, gspan,FFSM,grami}. These techniques have been widely applied in many domains to extract patterns~\cite{jiang2013survey}, such as chemistry, web, social network, and biology. Only a few works~\cite{AFSEM, FSMVLSI} have been done for VLSI gate-level netlist. Moreover, they mainly focused on directly applying FSM algorithms like ~\cite{gspan} on gate-level netlists to find frequent sub-circuits and the consideration of VLSI characteristics was absent in their solutions.
	
	\subsection{Challenges of Automated Standard Cell Library Extension}\label{challenges}
	
	To realize the co-optimization of designs and standard cell library and close the loop of netlist pattern mining and standard cell layout synthesis, we need to overcome the following challenges:

	\begin{itemize}
		\item Without identifying the promising pattern subgraphs as the inputs of standard cell layout synthesis, the resultant custom standard cell might not benefit the VLSI design. Meanwhile, simply selecting one or two patterns from the customization candidates to create new standard cells cannot realize the optimal solution, e.g., minimizing area cost. Therefore, we need to find the proper combination of patterns.
		\item Without the consideration of the requirements of VLSI design flow, FSM algorithms cannot find the optimal candidates for standard cell layout synthesis. For example, in previous works, their FSM solutions simply focused on the frequency of pattern recurrence in the gate-level netlist. However, pattern coverage and area saving after customization should be considered.  Moreover, the overlapped pattern subgraphs in the netlist should not be counted more than once since each node in the netlist can be mapped to only one standard cell during the mapping flow in logic synthesis. Finally, it is difficult to identify the patterns from large and complex gate-level netlist since the general FSM algorithms are timing-consuming and memory-intensive. 
	\end{itemize}
	
	With the consideration of the scenarios in real applications discussed above, the contributions of AutoCellLibX are highlighted as follows:
	
	\begin{itemize}
		\item A practical vertex encoding algorithm, which can find a proper set of neighbor vertex for pattern growth with the consideration of standard cell characteristics, as a part of high-efficiency FSM solution;
		\item A pattern growth algorithm that can expand the sizes of gate-level patterns while preserving their high recurrence frequencies. Compared to previous FSM approaches, our pattern growth solution carefully handle the overlaps between pattern subgraphs to meet the technology mapping constraint and maximize area reduction;
		\item A pattern combination algorithm which can iteratively find a set of gate-level patterns from numerous candidates as the extension part of the initial standard cell library to maximize the area reduction of the entire VLSI design; 
		\item To the best of our knowledge, it is the first automated standard cell extension framework that closes the optimization loop between the analysis of gate-level netlist and standard cell library customization. AutoCellLibX can generate SPICE netlists and GDSII layouts of the custom standard cells for downstream VLSI design flow.
	\end{itemize}

	\section{Preliminaries}\label{Preliminaries}
	
	In this section, we formulate the pattern mining problem in the scenarios of standard cell library extension and present an overview of our proposed framework.
	
	\subsection{Problem Formulation}\label{formulation}
	
	The post-technology mapping gate-level netlist can be formulated as a directed graph $G = (V, E, Lv, Le)$  with the following definitions:
	\begin{itemize}
		\item vertices $V(G) = \{v_1 , v_2, \dots , v_n \}$ represent cells after technology mapping in the gate-level netlist with flip-flops removed. The reason for such cell removal is that sequential circuits consist of combinational logic gates and sequential memory elements, and our targets of customization are the combinational logic gates.
		\item edges $E(G) =\{e_1, e_2, \dots , e_m\}$ represent $m$ pin-to-pin nets according to the driver-sink relationships in the gate-level netlist
		\item $L{v_i}$ represents the label for vertex $v_i$, which is the standard cell type of  $v_i$ 
		\item $L{e_i}$ represents  the label for edge $e_i$, which is a tuple $({Output}_i, {Input}_i)$, indicating that the edge $e_i$ connects an output pin named ${Output}_i$ of its driver cell  and an input pin named ${Input}_i$ of its sink cell.
		\item A pattern group $PGS_i = \{G'_1, G'_2, \dots , G'_{Ns} \}$ is a set of $N_s$ subgraphs of $G$. All the vertices in each subgraph in $PGS_i$ are connected.  The subgraphs in $PGS_i$ are isomorphic, and it means that the sub-circuits corresponding to these subgraphs have the same circuit functionality. They are not overlapped with each other, i.e., when  $  p\neq q$ and $ G'_p, G'_q\in PGS_i $, we have $V(G'_p) \cap V(G'_q)=\emptyset  $
		\item The coverage of a pattern group $Cov(PGS_i)$ is the number of vertices in it, i.e., $Cov(PGS_i) = | V(G'_1) \cup V(G'_2) \cup \dots \cup V(G'_{Ns})  |$, where $G'_j \in PGS_i$ \label{Coverage}
		\item A combination of pattern candidates $C = \{PGS_1 , PGS_2, \dots , PGS_{Np} \}$ is a set of ${N_p}$ pattern groups. The subgraphs in different pattern groups are not isomorphic, and they are not overlapped with each other, i.e., $\forall PGS_i,PGS_j \in C$, when $i\neq j$, $\forall G' \in PGS_i$, $\forall H' \in PGS_j$, $G'$ is not overlapped with $H'$.
		\item A reward function $R(C,G)$ indicates how much area can be saved for $G$ when  each subgraph $G'_i$ in the pattern groups of $C$ is replaced by a corresponding custom standard cell generated by standard cell layout synthesis flow.
	\end{itemize}
	
	According to the definitions above, the problem of standard cell library extension can be formulated as follows: given a gate-level netlist  $G$ and the initial standard cell library $LIB$,
	\begin{equation}
		\begin{aligned}
			\max_{C} \quad &  R(C,G)\\
			\textrm{s.t.} \quad & |C|<N_p    \\
			& \forall G'\in PGS_i\in C,  |G'|<S_p
		\end{aligned}
	\end{equation}
	where $N_p$ is the maximum number of custom standard cell types and $S_p$ is the maximum number of vertices in each pattern subgraph. These two parameters are determined by designers to limit the size of the extended library and the size of the custom standard cell. Empirically, in our implementation, $N_p$ and $S_p$ are set to be 5 and 10 respectively, since when these parameters become larger, the area benefit is marginal for the benchmarks. More discussion about these parameters can be found in Section~\ref{TerminationCriteriaofPatternMining} The resultant $C$ will be used to complement $LIB$ and generate a new extended $LIB'$.  
	
	\begin{figure}[b]
		\centering
		\includegraphics[width=\linewidth]{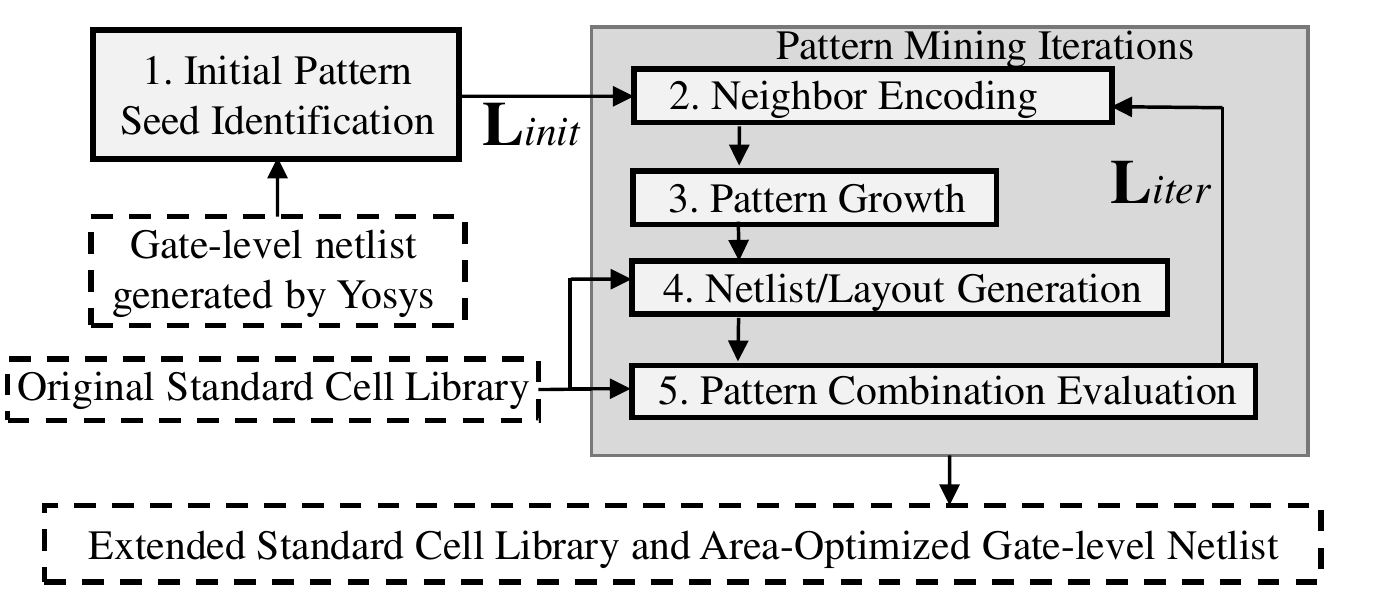}
		\caption{The outline of AutoCellLibX's workflow with 5 process phases } 
		\label{outline}
	\end{figure}

	\subsection{The Framework of AutoCellLibX}
	
	The workflow of AutoCellLibX is shown in Fig.~\ref{outline}. The inputs of AutoCellLibX are the post-technology mapping gate-level netlist (BLIF file) of the target VLSI design generated by Yosys~\cite{yosys} and an initial standard cell library used during the logic synthesis. The outputs are the SPICE netlists, GDSII layouts and the area reduction report of the combination of patterns. The workflow overview of AutoCellLibX is described in this subsection.
	
	\subsubsection{Initial Pattern Seed Identification} Initial pattern seed identification generates a set of initial 2-level tree-based patterns, $\mathbf{L}_{init}$, which will be the starting point of later iterative pattern growth for standard cell customization.
	
	\subsubsection{Neighbor Encoding} In the iterative loop, the first stage is to efficiently enumerate and encode the neighbor vertices of subgraphs in the target pattern group $PGS_{tgt}$ with the highest coverage in a list of sorted patterns, $\mathbf{L}_{iter}$ (or $\mathbf{L}_{init}$), to find the promising direction of the pattern growth.
	
	\subsubsection{Pattern Growth} Inspired by ~\cite{gspan}, some of the encoded neighbor vertices will be absorbed into the subgraphs in $PGS_{tgt}$ and  a new pattern group $PGS_{tgt}'$ with subgraphs of larger size is generated. Meanwhile, the global pattern information will be updated to improve the efficiency of later iterations of pattern mining.
	
	\subsubsection{Generation of SPICE Netlist and GDSII Layout}  According to the subgraph topology and label information in the new pattern group $PGS_{tgt}'$, a SPICE netlist indicating the interconnection of sized transistors for a potential custom standard cell will be generated. With the SPICE netlist and predefined design rules of the initial standard cell library, ASTRAN~\cite{ASTRAN},  the open-source standard cell layout synthesis tool, will be called to generate the GDSII layout of the custom cell.
	
	\subsubsection{Pattern Combination Evaluation} According to the generated GDSII layouts and the recurrence frequencies of the pattern groups, the overall benefit of a potential combination of patterns are evaluated by replacing corresponding subgraphs in $G$ with the custom standard cells. In this stage, AutoCellLibX will analyze the improvement potential of the patterns and make the decision whether further pattern mining iterations should be terminated.

	As shown in Fig.~\ref{outline}, neighbor encoding, pattern growth, generation of SPICE netlist and GDSII layout, and pattern combination evaluation are the five successive stages in the iterative pattern mining loop.
	
	\section{Implementation of AutoCellLibX}\label{implementation}
	
	\subsection{Initial Pattern Seed Identification}\label{InitialPatternSeedIdentification}
	
	The pattern growth procedure is to let some frequent subgraphs gradually absorb their neighbor vertices and initial patterns are the seeds of pattern growth in FSM algorithms. 
	
	In conventional solutions, like ~\cite{gspan,FFSM,grami}, which mainly count the frequencies of various subgraph patterns without other constraints, the initial patterns are simply all the edges with two interconnected vertices in $G$. For the general scenarios of standard cell customization without considering logic duplication, each vertex in the gate-level netlist should be mapped to only one standard cell during technology mapping. Edge-based initial patterns might lead to the result that some vertices will be covered by multiple pattern subgraphs, such as high-fanin cells and broadcast signals. For example, the NOR3X1 cell with index 0 is connected with two AND2X2 cells in Fig.~\ref{treePattern} but the NOR3X1 cell  can be mapped to only one custom standard cell with one of the two AND2X2 cells.
	
	To consider the constraint aforementioned, our pattern seed identification employs tree encoding as illustrated in Algorithm~\ref{initialPatternAlgorithm}. 
	
	\subsubsection{Initial Tree Encoding} In $G$, each vertex will be regarded as a root, and its 1-hop predecessors will construct a 2-level tree with the root vertex. Each of these trees will be encoded with a tree code, according to the steps shown in line 1-13 of Algorithm~\ref{initialPatternAlgorithm}. The tree code for a tree consists of two parts: the code of root (root code) and the code of leaves (leaf codes). The root code is simply the standard cell type of the root, like "[NOR3X1]". The leaf code for each leaf in the tree will be  a code including its standard cell type, and the label of the edge interconnecting the root and it, i.e., $Le_i$ in Section~\ref{formulation}. For example, "(AND2X2,Y,A)" for AND2X2 cell with index 1 in Fig.~\ref{treePattern} means that the output pin "Y" of an AND2X2 cell is connected to the input pin "A" of the root NOR3X1 cell. The tree code is the root code followed by the leaf codes of leaves, where the leaf codes are concatenated in the lexicographical order (line 5 in Algorithm~\ref{initialPatternAlgorithm}). For example, the tree code for the initial tree in red dash curve in Fig.~\ref{treePattern} is "[NOR3X1](AND2X2,Y,A)(AND2X2,Y,B)(NOR3X1,Y,C)".  
	
	It can be noticed that the tree code includes all the information of the corresponding tree, e.g., the vertices, the edges between them, and related labels. Meanwhile, due to the lexicographical order of the codes of leaves in tree code, each of these initial tree subgraphs will be mapped to only one tree code. Therefore, trees with the same tree code are isomorphic.
	
	As described in lines 8-13 in Algorithm~\ref{initialPatternAlgorithm}, AutoCellLibX will enumerate the vertices in $G$, construct corresponding pattern trees, encode the trees and record the recurrences of the tree patterns based on the codes. According to the recurrence record, we can rank the initial patterns in descending order of their recurrence frequencies. 
	
	\subsubsection{Overlap Elimination}\label{OverlapElimination} 
\begin{figure*}[!t]
		\setlength{\abovecaptionskip}{0.cm}
		\setlength{\belowcaptionskip}{-0.5cm}
		\centering
		\subfigure[Details of an example pattern]{
			\begin{minipage}[b]{0.25\linewidth}
				\includegraphics[width=\linewidth]{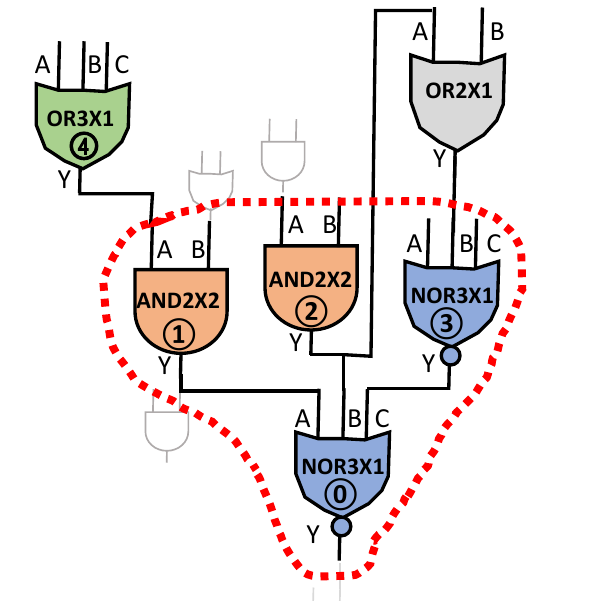} 
				\label{treePattern}
			\end{minipage}
		}
		\subfigure[Example of pattern overlap ]{
			\begin{minipage}[b]{0.20\linewidth}
				\includegraphics[width=\linewidth]{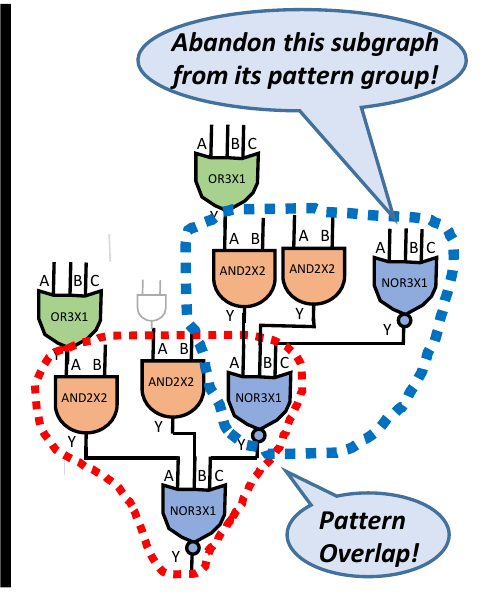} 
				\label{patternOverlap}
			\end{minipage}
		}
		\subfigure[Example of pattern recurrences: subgraphs in red dash curve are 5 separate isomorphic subgraphs in 1 $PGS_{tgt}$ ]{
			\begin{minipage}[b]{0.50\linewidth}
				\includegraphics[width=\linewidth]{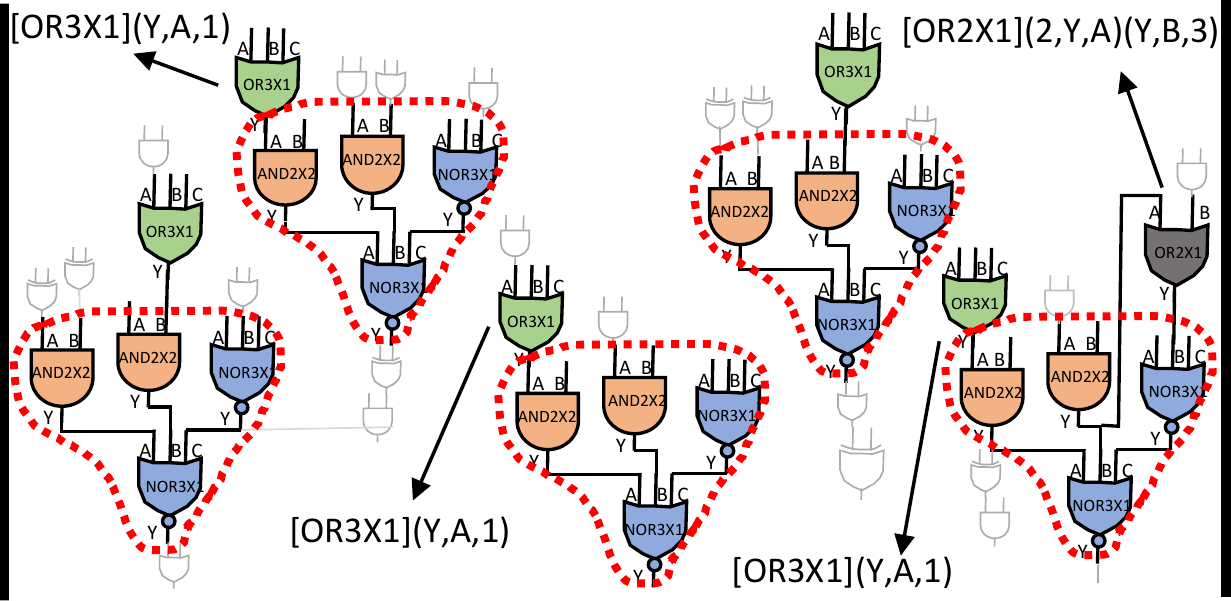} 
				\label{patternForest}
			\end{minipage}
		}
		
		\caption{Example for the pattern mining procedure: initial pattern seed identification, neighbor encoding and pattern growth.} 
		\label{patternMining}
	\end{figure*}

Please note that at this stage,  the pattern subgraphs could overlap with some other patterns at some of the vertices. As described in lines 14-21 in Algorithm~\ref{initialPatternAlgorithm}, to eliminate these overlaps, AutoCellLibX will enumerate the patterns from those most frequent ones to those less frequent to conduct checking. If all the vertices in a subgraph have not been occupied, these vertices will be marked as occupied. Otherwise, the subgraph will be abandoned from the record. As an example shown in Fig.~\ref{patternOverlap}, if two pattern subgraphs are overlapped, one of the subgraphs will be abandoned if it is visited later or belong to a  pattern with a lower frequency. In this way, AutoCellLibX removes the overlaps among the subgraphs in the patterns and preserves those high-frequency subgraph patterns for the reuse of custom standard cells. After the elimination of the overlaps, the pattern groups with different tree codes will be recorded in $\mathbf{L}_{init}$ and ranked in descending order of their coverage in $G$, i.e. $Cov(PGS_i)$ defined in Section~\ref{Coverage}, for later iterative pattern growth, since it is more likely for the patterns with high coverage to realize a noticeable impact on the entire design according to Section~\ref{BenefitEvaluationFunction}.

	\begin{algorithm}[t]
		\small
		\DontPrintSemicolon
		\KwIn{the gate-level netlist after technology mapping $G$}
		\KwOut{A pattern list in descending order of the frequencies of the initial patterns $\mathbf{L}_{init}=\{{PGS}_i\}$}
		
		\SetKwFunction{FMain}{TreeEncode}
		\SetKwProg{Fn}{Function}{:}{}
		\Fn{\FMain{root, predecessors}}{
			predCodes = \{\};
			
			\ForEach{$v_i$ $\in$ predecessors}
			{
				predCodes.append(stdCellType($v_i$)+$Le$(root,$v_i$));
			}
			\textbf{return}  concat(stdCellType(root), lexSort(predCodes));
		}

		code2Trees = record of the trees with the same code;
		
		code2Freq = record of the code frequency;
		
		\tcc{construct trees and encode them}

		\ForEach{root $\in$ $V$} 
		{

			preds = getPreds(root); // get direct 1-hop predecessors
			
			treeCode = TreeEncode(root,preds);
			
			code2Freq[treeCode] += 1;
			
			code2Trees[treeCode].append(getTree(root));
			
		}
		
		sortedCodes = sortPatternCodes(code2Freq);
		
		\tcc{eliminate overlaps}
		
		\ForEach{code $\in$ sortedCodes} 
		{
			\ForEach{tree $\in$ code2Trees[code]} 
			{		
				
				\If{allVerticesAreUnoccupied(tree)}
				{
					OccupyAndIndexAllVerticesIn(tree);
				}
				\Else
				{
					code2Freq[code] -= 1;
					
					code2Trees[code].remove(tree);
				}
				
			}
		}
		
		$\mathbf{L}_{init}$ = sortedPatternListAccordingToCoverage(code2Trees);
		
		\Return{$\mathbf{L}_{init}$}

		\caption{Initial Pattern Seed Identification}
		\label{initialPatternAlgorithm}
	\end{algorithm}

	\subsubsection{Post-process for Iterative Pattern Mining} Each vertex in a tree will be assigned an index in the tree to facilitate isomorphism checking among subgraphs with the same pattern in later procedures. Root vertex will have an index of 0 while the predecessor vertices will be indexed from 1 to the number of predecessors, according to the lexicographical order of their leaf codes. For example, as shown in Fig.~\ref{treePattern}, the bottom NOR3X1 cell has an index of 0, and the two AND2X2 cells are assigned index 1 and 2 respectively. This index will be used in neighbor encoding in Section~\ref{NeighborEncodingBasedonInterconnectionFeatures}.

	\begin{algorithm}[!t]
		\small
		\DontPrintSemicolon
		\KwIn{the top-1 pattern group $PGS_{tgt}$ from $\mathbf{L}_{init}$ or $\mathbf{L}_{iter}$ }
		\KwOut{the record of the neighbor vertices of $PGS_{tgt}$ and their pattern codes}
		
		\SetKwFunction{FMain}{NeighborEncode}
		\SetKwProg{Fn}{Function}{:}{}
		\Fn{\FMain{neighbor, patternSubG}}{
			codes = \{\};
			
			\tcc{ get the edge codes}
			
			\ForEach{$v_i$ $\in$ patternSubG}
			{
				viIndex = getIndex($v_i$, patternSubG);
				
				\If {neighbor is driver of $v_i$}
				{
					codes.append($Le$(neighbor,$v_i$,)+viIndex);
				}
				\Else{codes.append(viIndex+$Le$( $v_i$, neighbor));}
				
			}
			\textbf{return}  concat(stdCellType(neighbor), lexSort(codes));
		}

		code2Neighbors=record of the neighbors with the same code;
		
		neighbor2Subgraph = record of the owner subgraph of a neighbor;
		
		code2Freq = record of the code frequency;
		
		encoded = record of encoded vertices;

		\tcc{enumerate the neighbors of the subgraphs in $PGS_{tgt}$ and encode them}

		\ForEach{subG  $\in PGS_{tgt}$} 
		{
			\ForEach {neighbor $\in$  1-hop neighbors of subG} 
			{
				\If {NOT encoded[neighbor]}
				{
					code = NeighborEncode(neighbor, subG);
					
					neighbor2Subgraph[neighbor] = subG;
					
					encoded[neighbor] = True; // avoid overlaps				
					
					code2Neighbors[code].append(neighbor);
				}
			}
			
		}

		\Return{code2Neighbors, neighbor2Subgraph}

		\caption{Neighbor Encoding}
		\label{neighborEncodingAlgorithm}
	\end{algorithm}
	
	\subsection{Neighbor Encoding}\label{NeighborEncoding}

	Our pattern mining procedure is based on pattern growth which gradually merges some selected vertices into the subgraphs of existing patterns to enlarge the pattern size while trying to preserve the recurrence frequencies of the patterns. Therefore, selecting proper neighbors is a critical part of pattern mining, which will be illustrated in this subsection.
	
	AutoCellLibX will take the top-1 pattern group $PGS_{tgt}$ which has the highest coverage in the gate-level netlist, from $\mathbf{L}_{init}$ or $\mathbf{L}_{iter}$, as the input of neighbor encoding as depicted in Fig.~\ref{outline}. Correspondingly, the outputs of neighbor encoding are a set of neighbor vertices of $PGS_{tgt}$ and their pattern codes. The overall steps of neighbor encoding are presented in Algorithm~\ref{neighborEncodingAlgorithm}.
	
	\subsubsection{Neighbor Enumeration}  As shown in lines 14-20 of Algorithm~\ref{neighborEncodingAlgorithm}, in this stage, AutoCellLibX will enumerate the neighbors of all the subgraphs in $PGS_{tgt}$ to conduct encoding. Here, a neighbor of a subgraph is defined as a vertex that is not a vertex in the subgraph but is a 1-hop neighbor (predecessor or successor) of a vertex in the subgraph. The output of NeiborEncode function is determined by both the neighbor and the subgraph. A neighbor might be shared by several subgraphs of the same pattern but it will be encoded once when it is visited for the first time, as indicated by line 16 of Algorithm~\ref{neighborEncodingAlgorithm}. This constraint can ensure that there will be no overlap among the later extended pattern subgraphs and AutoCellLibX will record the owner subgraph of each neighbor as shown in line 18 of Algorithm~\ref{neighborEncodingAlgorithm}.

	\subsubsection{Neighbor Encoding with Interconnection Features}\label{NeighborEncodingBasedonInterconnectionFeatures} The code of a neighbor (neighbor code) consists of its standard cell type, and the labels of edges between the neighbor and the subgraph, as demonstrated in line 3-9 of Algorithm~\ref{neighborEncodingAlgorithm}. Each edge between the neighbor and the subgraph will be enumerated and a corresponding edge code of the edge will be generated. As mentioned in Section~\ref{InitialPatternSeedIdentification} and ~\ref{PatternGrowth}, each vertex in an existing subgraph of a specific pattern has its index in the subgraph. Accordingly, the edge code includes the index of the vertex in the subgraph and the label of edges. The order of these two parts in the edge code depends on whether the neighbor is the driver of the edge.  For example, if the subgraph in the red dash curve is the target subgraph, the edge between OR2X1 cell and NOR3X1 in Fig.~\ref{treePattern} will be encoded as "(Y,B,3)" while the edge between OR2X1 cell and AND2X2 in Fig.~\ref{treePattern} will be encoded as "(2,Y,A)".  The neighbor code is constructed as the standard cell type of the neighbor vertex followed by the edge codes of edges between the neighbor vertex and the corresponding subgraph, where the edge codes are concatenated in the lexicographical order. For example, as shown in Fig.~\ref{treePattern}, the neighbor code of the OR2X1 cell will be "[OR2X1](2,Y,A)(Y,B,3)" if the subgraph in the red dash curve is the target  subgraph for neighbor encoding. Consequently, the recurrence frequencies of the OR2X1 cell (encoded as "[OR2X1](2,Y,A)(Y,B,3)") and OR3X1 cell (encoded as "[OR3X1](Y,A,1)") is 1 and 3 respectively  in Fig.~\ref{patternForest}.
	
	Please be aware that some input pins of some standard cells, e.g., the input pin "A" and "B" of OR2X1 cell in Fig.~\ref{treePattern}, have equivalent functionality and highly similar parasitics. Switching the interconnection edges among these pins will not change the logic functionality. For the sake of this factor, during the actual implementation of edge code generation, equivalent pins will be encoded to the same name. For example, pin "B" of OR2X1 will be encoded as "A".
	
	Since the neighbor code encodes all the necessary information of the interconnection between the neighbor and the vertices in the subgraph in a unique form, those neighbors with the same neighbor code have the isomorphic interconnection egdes with their corresponding subgraphs. Therefore, the neighbor codes can classify the neighbor vertices into different sets for later pattern growth.
	
	In our proposed solution, the time complexity of neighbor encoding is O($|E|$), determined by lines 14-20 of Algorithm~\ref{neighborEncodingAlgorithm}, where neighbor vertices of the subgraphs are enumerated for encoding.

	\newtheorem{mytheorem}{Theorem}
	
	\subsection{Pattern Growth}\label{PatternGrowth}
	
	In the stage of pattern growth, AutoCellLibX merges some neighbor vertices into the subgraphs of $PGS_{tgt}$ to enlarge the pattern size while trying to preserve the recurrence frequencies of the original pattern to realize a high coverage of the new pattern group.

	\subsubsection{Fundamentals of The Selection of Neighbors} Selecting neighbors for pattern growth is based on two theorems verified in classic subgraph mining solutions~\cite{apriori, gspan}.
	
	\begin{mytheorem}\label{absorbVertex}
		Given  two isomorphic subgraph $G_A(V_A,E_A)$ and $G_B(V_B,E_B)$ of $G$, and they have neighbor vertex $v_a$ and $v_b$ respectively. If $v_a$ and $v_b$ have   isomorphic interconnection edge set $E_a$ and $E_b$ connected to $G_A$ and $G_B$ respectively (i.e., $v_a$ and $v_b$  have the same neighbor codes),  $G'_A(V'_A,E'_A)$ and $G'_B(V'_B,E'_B)$ are isomorphic, where $V'_A = v_a \cup  V_a, V'_B = v_b \cup  V_b, E'_A = E_a \cup  E_A, E'_B = E_b \cup  E_B$  ~\cite{gspan}.
	\end{mytheorem}
	
	\begin{mytheorem}\label{frequencyDescending}
		(Recurrence Frequency Monotonicity): If a subgraph pattern $G_s$ recurs $R$ times in $G$, then any subgraph of $G_s$ recurs no less than $R$ time in $G$ ~\cite{apriori}.
	\end{mytheorem}
	
	Based on Theorem~\ref{absorbVertex}, we can merge those neighbor vertices with the same neighbor code, as well as the interconnection edges between them and their owner subgraphs, into the owner subgraphs in $PGS_{tgt}$ to obtain a new pattern group $PGS_{new}$, where isomorphic subgraphs get larger compared to those in $PGS_{tgt}$. However, the size of $PGS_{new}$ will not be greater than the size of $PGS_{tgt}$, since some of the subgraphs in $PGS_{tgt}$ do not have a neighbor with the specific neighbor code. This result is supported by Theorem~\ref{frequencyDescending}. 
	
	As the example shown in Fig.~\ref{patternForest}, suppose the subgraphs in the red dash curve are those in $PGS_{tgt}$, the size of which is 5. Only 3 subgraphs have a neighbor OR3X1 cell encoded as "[OR3X1](Y,A,1)" and it means that if we merge these OR3X1 cells into the corresponding subgraphs in $PGS_{tgt}$, we will get a new pattern group $PGS_{new}$, the size of which will be 3.

	\begin{algorithm}[!t]
		\small
		\DontPrintSemicolon
		\KwIn{$\mathbf{L}_{iter}$ (or $\mathbf{L}_{init}$), the top-1 pattern group $PGS_{tgt}$, code2Neighbors, neighbor2Subgraph in Algorithm~\ref{neighborEncodingAlgorithm} }
		\KwOut{updated list of pattern groups $\mathbf{L'}_{iter}$}

		targetCode = getFreqNeighhborCode(code2Neighbors);
		
		targetNeighbors = code2Neighbors[targetCode];
		
		$PGS_{new}$ = \{\}

		\tcc{enumerate the neighbors with the target neighbor code}

		\ForEach{$v_i$ $\in$ targetNeighbors} 
		{
			ownerSubG = neighbor2Subgraph[neighbor];
			
			\If {ownerSubG is NOT abandoned}
			{
				\tcc{deal with neighbor occupied by other pattern subgraph }
				\If {$\exists PGS'$, $v_i$ $\in$ $PGS'$ $\in \mathbf{L}_{iter}$}
				{
					overlapSubG = getSubgraphContain($v_i$);
					
					overlapSubG.setAbandoned();
					
					$PGS'$.remove(overlapSubG);
				}
				
				setIndexInSubGragh($v_i$,ownerSubG);
				
				newSubG = ownerSubG $\cup$ $v_i$ $\cup$  E($v_i$,ownerSubG);
				
				ownerSubG.setAbandoned();
				
				$PGS_{tgt}$.remove(ownerSubG);
				
				$PGS_{new}$.append(newSubG);

			}
		}

		$\mathbf{L}_{supSeed}$ = patternSeedSupplement($G$);
		
		$\mathbf{L}'_{iter}$ = sortByCoverage($\mathbf{L}_{iter}$+$PGS_{new}$+$L_{supSeed}$);
		
		removeLowCoveragePatternGroups($\mathbf{L}'_{iter}$);
		
		\Return{$\mathbf{L}'_{iter}$}

		\caption{Pattern Growth}
		\label{patternGrowthAlgorithm}
	\end{algorithm}
	
	\subsubsection{Acquisition of New Pattern Group} 
	To realize high coverage of $PGS_{new}$, AutoCellLibX will select the neighbor code which has the most recurrences among the subgraphs in $PGS_{tgt}$, e.g., "[OR3X1](Y,A,1)" for the subgraphs  in Fig.~\ref{patternForest}, and merge the neighbors with the selected neighbor code into their owner subgraphs.  When merging a neighbor into a pattern subgraph, the vertex will be assigned an index of the number of vertices of the original subgraph. This index indicates the "posistion" of the vertex in the pattern subgraph and  will be used in neighbor encoding in Section~\ref{NeighborEncodingBasedonInterconnectionFeatures}.
	
	Those extended subgraphs will be removed from $PGS_{tgt}$ and added to $PGS_{new}$. Meanwhile, similar to the example in Fig.~\ref{patternOverlap} for initial pattern seed generation, if those involved neighbors originally belong to the subgraphs in some pattern groups in $\mathbf{L}_{iter}$ (or $\mathbf{L}_{init}$), the subgraphs will be abandoned by the corresponding pattern groups and vertices in those subgraphs will be marked as unoccupied to meet the no-overlap constraint. These steps are demonstrated in lines 4-18 of Algorithm~\ref{patternGrowthAlgorithm}. In this procedure,  AutoCellLibX tries to raise the coverage of $PGS_{new}$ and shrink the low-coverage pattern groups. 
	
	\subsubsection{Post-process for Later Pattern Mining}\label{PostprocessforLaterPatternMining}
	Since some subgraphs are eliminated from their pattern groups due to overlaps during pattern growth, some vertices in $G$ become unoccupied. These vertices will be fed into a function for pattern seed supplement, which will generate the seed patterns with these unoccupied vertices, similar to initial pattern seed generation in Section~\ref{InitialPatternSeedIdentification}.
	
	The pattern groups in $\mathbf{L}'_{iter}$ will be sorted in descending order of their coverage in $G$ to maximize area reduction with standard cell customization. Details for the relationship between pattern coverage and area benefit are illustrated in Section~\ref{BenefitEvaluationFunction}. Meanwhile, pattern groups with coverage lower than 2.5\% of the number of  vertices in $G$ will be removed from $\mathbf{L}'_{iter}$ to prune unnecessary exploration.
	
	The time complexity of pattern growth is O($|V|$), determined by lines 4-15 of Algorithm~\ref{patternGrowthAlgorithm}, where neighbor vertices with the same neighbor code of the subgraphs are enumerated.

	\subsection{Generation of SPICE Netlist and GDSII Layout}\label{GenerationofSPICENetlistandGDSIILayout}
	
	AutoCellLibX needs to generate SPICE netlists and GDSII layouts for the patterns, so it can evaluate their area benefits. 
	
	\subsubsection{Selection of Patterns for Generation} \label{SelectionofPatternsforGeneration}
	While there are many pattern groups in $\mathbf{L}_{iter}$ and standard cell layout synthesis is time-consuming, just a few pattern groups should be considered for customization. According to the empirical observation on our benchmarks, the pattern groups outside the top-5 in $\mathbf{L}_{iter}$ cover less than 10\% of vertices in $G$ during the runtime of AutoCellLibX. Therefore, AutoCellLibX will only generate the SPICE netlist and GDSII layout for the top-$N_p$ pattern groups in $\mathbf{L}_{iter}$ for each mining iteration, where $N_p$ is the maximum number of custom standard cell types defined in Section~\ref{Preliminaries}.
	
	\subsubsection{SPICE Netlist Generation}
	As mentioned in Section~\ref{Preliminaries}, $G$ is a gate-level netlist after technology mapping, where each vertex is actually a standard cell. In the standard cell library, the SPICE netlist for each cell is provided. For standard cell merging, a pattern subgraph represents a set of interconnected standard cells. Therefore, AutoCellLibX can merge the SPICE netlists of the cells by connecting the pins of sized transistors according to the topology of the subgraph. To make the most of standard cell customization, AutoCellLibX will check all the I/O signals in the pattern subgraph, and if an I/O signal has no interconnection outside the subgraph in $G$ and becomes a completely internal signal of the pattern subgraph, the I/O pin for this signal on the SPICE interface of this custom standard cell will be removed. This can lower the pressure of I/O pin placement and provide more flexibility for the other parts of placement and routing during layout synthesis.
	
	\subsubsection{GDSII Layout Generation}
	Based on the SPICE netlist and the design rules of the standard cell library (e.g., number of routing tracks), AutoCellLibX will involve ASTRAN~\cite{ASTRAN} to conduct layout synthesis for a specified pattern subgraph.

	\subsection{Pattern Combination Evaluation}\label{PatternCombinationEvaluation}
	
	After previous stages of the iterative mining loop, at this stage, AutoLibCellX obtains a list of sorted pattern groups which do not overlap with each other and the area benefit of a combination of pattern groups should be evaluated.
	
	\subsubsection{Reward Function}\label{BenefitEvaluationFunction} The overall reward function $R(C,G)$ indicating how much area could be saved can be formulated as:	
	\begin{eqnarray}
		R(C,G)=\sum_{PGS_i \in C} [|PGS_i| * F(PatternG_i)]
	\end{eqnarray}	
	where $PGS_i$ are the pattern groups in pattern combination $C$, $PatternG_i$ is one of the subgraphs in $PGS_i$, and $F$ is a function evaluating the area benefit of the custom standard cell generated according to $PatternG_i$ compared to the combination of original standard cells.
	
	\begin{figure}[b]
		\centering
		\includegraphics[width=0.8\linewidth]{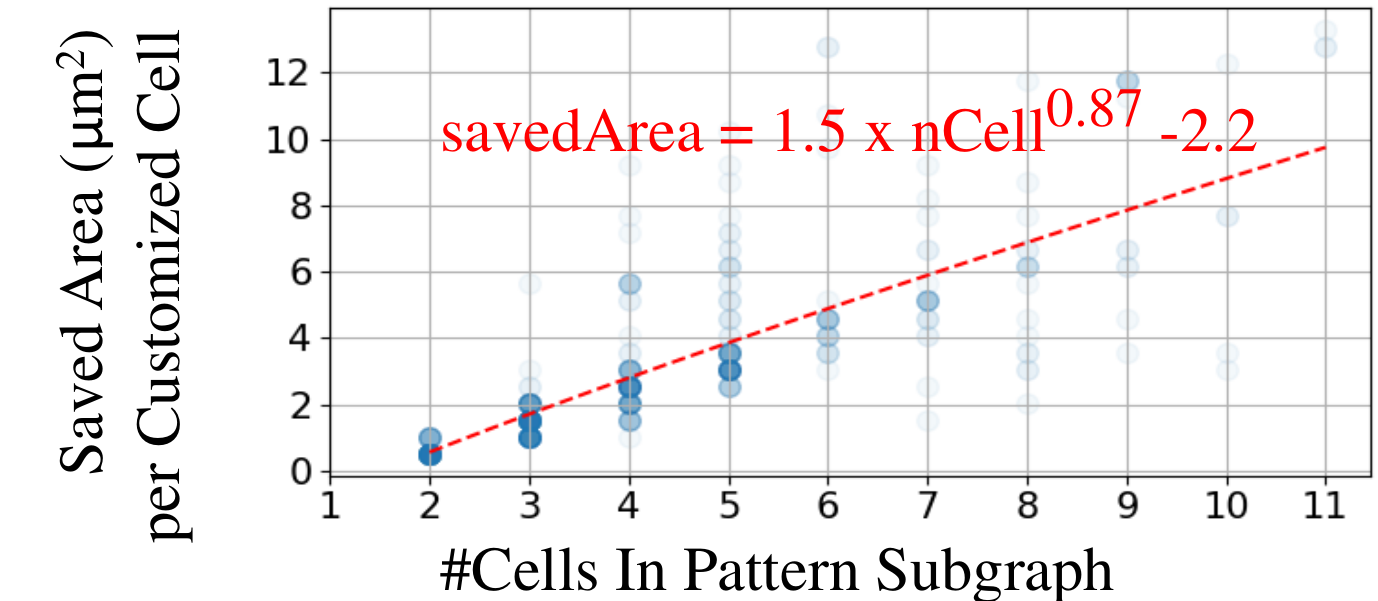}
		\caption{Example of 50 custom standard cells showing the correlation between the pattern size and the  area saved with the custom cell for the pattern} 
		\label{patternSize}
	\end{figure}
	
	To find out the factors impacting $ F(PatternG_i)$, a series of experiments have been conducted and the empirical results of 50 custom standard cells generated by ASTRAN~\cite{ASTRAN} based on FreePDK45 process~\cite{freepdk45} are shown in Fig.~\ref{patternSize}. According to the results, the area benefit of a custom standard cell is nearly proportional to $PatternSize_i$, the number of original standard cells in the pattern subgraph $PatternG_i$. Therefore, we can get $R_{approx}(C,G)$, an approximation of $R(C,G)$ as follows:
	\begin{eqnarray}
		R(C,G)&&\approx K\sum_{PGS_i \in C} [|PGS_i| * PatternSize_i] \\
		&& = K\sum_{PGS_i \in C} Cov(PGS_i) = R_{approx}(C,G)
	\end{eqnarray}	 where $K$ is a linear approximation factor. Accordingly, for a specific iteration, selecting the top-$N_p$ pattern groups from $\mathbf{L}_{iter}$, which is sorted according to pattern coverage (i.e., $R_{approx}(C,G)$), as the pattern combination $C$ to generate custom standard cells is the near-optimal solution from the perspective of area benefit. Such a relationship between pattern coverage and area benefit explains why AutoCellLibX sorts the pattern groups by coverage in Section~\ref{PostprocessforLaterPatternMining} and ~\ref{OverlapElimination}.

	\subsubsection{Termination Criteria of Pattern Mining}\label{TerminationCriteriaofPatternMining}  The area benefit of the top-$N_p$ pattern groups could change as the pattern mining loop iterates. Therefore, AutoCellLibX should stop the loop at a proper iteration to reach the optimal combination of custom standard cells.
	
	According to Theorem~\ref{frequencyDescending} (recurrence frequency monotonicity), it can be noticed that the values of $|PGS_i|$ tend to have a descending trend, while the values of $PatternSize_i$ are increasing, during the iterative pattern mining. As indicated in Section~\ref{Preliminaries}, $PatternSize_i$ will be limited within $S_p$ so $R(C,G)$ has an upper bound for each benchmark. Moreover, the values of $PatternSize_i$, starting from 1, increase by at most 1 in each iteration while the declines of $|PGS_i|$ are usually greater than 20\% in each iteration for the benchmarks in our experiments. It means that when $PatternSize_i$ is greater than 5,  it is much harder for the increase of $PatternSize_i$ to compensate for the decline of $|PGS_i|$.
	
	Based on these analyses, AutoCellLibX will terminate the pattern mining loop if $R(C,G)$ declines in the last two iterations since it has little probability to realize higher area benefit if the loop continues.

	\section{Experimental Results}
	
	In this section, experiments to comprehensively evaluate the performance of AutoCellLibX are presented.
	
	\subsection{Target Process, Benchmarks and Environment}\label{experimentSetting}
	
	Our evaluation experiments are based on FreePDK45~\cite{freepdk45}, an open-source generic process design kit developed by Oklahoma State University and North Carolina State University, which uses a predictive 45nm CMOS technology process. Concretely, in the experiments, the benchmarks are synthesized by Yosys~\cite{yosys} based on the FreePDK45 technology library, SPICE netlists of custom standard cells are obtained according to the SPICE netlists of FreePDK45 standard cells, and ASTRAN generates the layout of custom standard cells according to the design rules of FreePDK45.
	
	To comprehensively evaluate AutoCellLibX in different scenarios, we collect 31 open-source benchmarks from various domains, including the EPFL combinational benchmark suite~\cite{epfl}, BOOM~\cite{boom}, a high-performance RISC-V CPU design, and Gemmini~\cite{gemmini}, a systolic array accelerator design for deep learning. For larger designs like BOOM and Gemmini which include multiple submodules with significantly different functionality and logic characteristics, e.g., DCache and ALU in BOOM, it is impractical for AutoCellLibX to find common patterns among these unique submodules. AutoCellLibX provides a hierarchical partitioning tool that can partition the design to a given depth so designers can select some of the partitions and find the  custom standard cells for each of them. As mentioned in Section~\ref{formulation}, $N_p$ and $S_p$ are set to be 5 and 10 respectively.
	
	Experiments are conducted on Ubuntu 20.04 with Intel i7-9850H CPU (2.60 GHz, 12 logic cores) and 32GB DDR4. 
	
\begin{table}[!t]
	\caption{Experimental Results}
	\centering
	\smaller
	\setlength\tabcolsep{2pt} 
\begin{tabular}{ccccclcc}
	\hline
	\multicolumn{1}{|c|}{Benchmark}  & \multicolumn{1}{c|}{\begin{tabular}[c]{@{}c@{}}Original\\ Area(um2)\end{tabular}} & \multicolumn{1}{c|}{\begin{tabular}[c]{@{}c@{}}Optimized\\Area(um2)\end{tabular}} & \multicolumn{1}{c|}{\begin{tabular}[c]{@{}c@{}}Reduce\\Area(\%)\end{tabular}} & \multicolumn{1}{c|}{\begin{tabular}[c]{@{}c@{}}Run\\-time(s)\end{tabular}} & \multicolumn{1}{l|}{\begin{tabular}[c]{@{}c@{}}FSM\\Time(s)\end{tabular}} & \multicolumn{1}{c|}{\begin{tabular}[c]{@{}c@{}}Pattern\\ Sizes\end{tabular}} & \multicolumn{1}{c|}{\begin{tabular}[c]{@{}c@{}}Pattern\\Cov(\%)\end{tabular}} \\ \hline
\multicolumn{8}{c}{EPFL Benchmark Suite~\cite{epfl}}                                                                                                                                                                                                                                                                                                                                                                                                                                                                                                                                                                                \\ \hline
\multicolumn{1}{|c|}{adder}      & \multicolumn{1}{c|}{1890}                                                          & \multicolumn{1}{c|}{1654}                                                           & \multicolumn{1}{c|}{12.51}                                                      & \multicolumn{1}{c|}{1314}                                                  & \multicolumn{1}{c|}{0.02}                                                  & \multicolumn{1}{c|}{4/3/-/-/-}                                               & \multicolumn{1}{c|}{56.86}                                                     \\ \hline
\multicolumn{1}{|c|}{priority}   & \multicolumn{1}{c|}{1111}                                                          & \multicolumn{1}{c|}{983}                                                            & \multicolumn{1}{c|}{11.53}                                                      & \multicolumn{1}{c|}{475}                                                   & \multicolumn{1}{c|}{0.02}                                                  & \multicolumn{1}{c|}{3/3/4/-/-}                                               & \multicolumn{1}{c|}{39.22}                                                     \\ \hline
\multicolumn{1}{|c|}{multiplier} & \multicolumn{1}{c|}{40326}                                                         & \multicolumn{1}{c|}{35976}                                                          & \multicolumn{1}{c|}{10.79}                                                      & \multicolumn{1}{c|}{1554}                                                  & \multicolumn{1}{c|}{0.24}                                                  & \multicolumn{1}{c|}{5/3/3/4/2}                                               & \multicolumn{1}{c|}{47.34}                                                     \\ \hline
\multicolumn{1}{|c|}{div}        & \multicolumn{1}{c|}{45283}                                                         & \multicolumn{1}{c|}{40669}                                                          & \multicolumn{1}{c|}{10.19}                                                      & \multicolumn{1}{c|}{2222}                                                  & \multicolumn{1}{c|}{0.38}                                                  & \multicolumn{1}{c|}{5/2/3/2/4}                                               & \multicolumn{1}{c|}{45.74}                                                     \\ \hline
\multicolumn{1}{|c|}{dec}        & \multicolumn{1}{c|}{630}                                                           & \multicolumn{1}{c|}{593}                                                            & \multicolumn{1}{c|}{5.85}                                                       & \multicolumn{1}{c|}{158}                                                   & \multicolumn{1}{c|}{0.02}                                                  & \multicolumn{1}{c|}{4/-/-/-/-}                                               & \multicolumn{1}{c|}{15.69}                                                     \\ \hline
\multicolumn{1}{|c|}{voter}      & \multicolumn{1}{c|}{17303}                                                         & \multicolumn{1}{c|}{16339}                                                          & \multicolumn{1}{c|}{5.57}                                                       & \multicolumn{1}{c|}{2407}                                                  & \multicolumn{1}{c|}{0.09}                                                  & \multicolumn{1}{c|}{2/3/3/3/3}                                               & \multicolumn{1}{c|}{35.33}                                                     \\ \hline
\multicolumn{1}{|c|}{square}     & \multicolumn{1}{c|}{31115}                                                         & \multicolumn{1}{c|}{29389}                                                          & \multicolumn{1}{c|}{5.55}                                                       & \multicolumn{1}{c|}{1228}                                                  & \multicolumn{1}{c|}{0.13}                                                  & \multicolumn{1}{c|}{3/4/3/3/3}                                               & \multicolumn{1}{c|}{26.98}                                                     \\ \hline
\multicolumn{1}{|c|}{max}        & \multicolumn{1}{c|}{4298}                                                          & \multicolumn{1}{c|}{4079}                                                           & \multicolumn{1}{c|}{5.1}                                                        & \multicolumn{1}{c|}{415}                                                   & \multicolumn{1}{c|}{0.03}                                                  & \multicolumn{1}{c|}{2/3/4/4/2}                                               & \multicolumn{1}{c|}{42.67}                                                     \\ \hline
\multicolumn{1}{|c|}{arbiter}    & \multicolumn{1}{c|}{16990}                                                         & \multicolumn{1}{c|}{16403}                                                          & \multicolumn{1}{c|}{3.45}                                                       & \multicolumn{1}{c|}{3323}                                                  & \multicolumn{1}{c|}{0.15}                                                  & \multicolumn{1}{c|}{7/2/4/4/2}                                               & \multicolumn{1}{c|}{15.29}                                                     \\ \hline
\multicolumn{1}{|c|}{cavlc}      & \multicolumn{1}{c|}{1086}                                                          & \multicolumn{1}{c|}{1055}                                                           & \multicolumn{1}{c|}{2.87}                                                       & \multicolumn{1}{c|}{266}                                                   & \multicolumn{1}{c|}{0.02}                                                  & \multicolumn{1}{c|}{4/2/3/-/-}                                               & \multicolumn{1}{c|}{11.64}                                                     \\ \hline
\multicolumn{1}{|c|}{sqrt}       & \multicolumn{1}{c|}{45259}                                                         & \multicolumn{1}{c|}{43996}                                                          & \multicolumn{1}{c|}{2.79}                                                       & \multicolumn{1}{c|}{3869}                                                  & \multicolumn{1}{c|}{0.25}                                                  & \multicolumn{1}{c|}{4/3/4/4/3}                                               & \multicolumn{1}{c|}{29.39}                                                     \\ \hline
\multicolumn{1}{|c|}{int2float}  & \multicolumn{1}{c|}{370}                                                           & \multicolumn{1}{c|}{360}                                                            & \multicolumn{1}{c|}{2.49}                                                       & \multicolumn{1}{c|}{344}                                                   & \multicolumn{1}{c|}{0.02}                                                  & \multicolumn{1}{c|}{3/4/-/-/-}                                               & \multicolumn{1}{c|}{13.42}                                                     \\ \hline
\multicolumn{1}{|c|}{router}     & \multicolumn{1}{c|}{332}                                                           & \multicolumn{1}{c|}{325}                                                            & \multicolumn{1}{c|}{2.31}                                                       & \multicolumn{1}{c|}{1031}                                                  & \multicolumn{1}{c|}{0.01}                                                  & \multicolumn{1}{c|}{4/-/-/-/-}                                               & \multicolumn{1}{c|}{8.7}                                                       \\ \hline
\multicolumn{1}{|c|}{sin}        & \multicolumn{1}{c|}{9155}                                                          & \multicolumn{1}{c|}{8976}                                                           & \multicolumn{1}{c|}{1.95}                                                       & \multicolumn{1}{c|}{707}                                                   & \multicolumn{1}{c|}{0.05}                                                  & \multicolumn{1}{c|}{5/3/-/-/-}                                               & \multicolumn{1}{c|}{7.55}                                                      \\ \hline
\multicolumn{1}{|c|}{mem\_ctrl}  & \multicolumn{1}{c|}{66114}                                                         & \multicolumn{1}{c|}{64881}                                                          & \multicolumn{1}{c|}{1.86}                                                       & \multicolumn{1}{c|}{475}                                                   & \multicolumn{1}{c|}{0.40}                                                  & \multicolumn{1}{c|}{2/2/3/3/3}                                               & \multicolumn{1}{c|}{9.89}                                                      \\ \hline
\multicolumn{1}{|c|}{i2c}        & \multicolumn{1}{c|}{1818}                                                          & \multicolumn{1}{c|}{1793}                                                           & \multicolumn{1}{c|}{1.38}                                                       & \multicolumn{1}{c|}{105}                                                   & \multicolumn{1}{c|}{0.01}                                                  & \multicolumn{1}{c|}{3/2/-/-/-}                                               & \multicolumn{1}{c|}{6.59}                                                      \\ \hline
\multicolumn{1}{|c|}{ctrl}       & \multicolumn{1}{c|}{195}                                                           & \multicolumn{1}{c|}{193}                                                            & \multicolumn{1}{c|}{0.79}                                                       & \multicolumn{1}{c|}{22}                                                    & \multicolumn{1}{c|}{0.01}                                                  & \multicolumn{1}{c|}{2/-/-/-/-}                                               & \multicolumn{1}{c|}{7.41}                                                      \\ \hline
\multicolumn{1}{|c|}{Average}    & \multicolumn{2}{c|}{}                                                                                                                                                    & \multicolumn{1}{c|}{5.11}                                                       & \multicolumn{1}{c|}{1171}                                                  & \multicolumn{1}{c|}{0.11}                                                  & \multicolumn{1}{c|}{}                                                        & \multicolumn{1}{c|}{24.69}                                                     \\ \hline
\multicolumn{8}{c}{BOOM~\cite{boom}}                                                                                                                                                                                                                                                                                                                                                                                                                                                                                                                                                                                                \\ \hline
\multicolumn{1}{|c|}{PTW}        & \multicolumn{1}{c|}{394413}                                                        & \multicolumn{1}{c|}{375985}                                                         & \multicolumn{1}{c|}{4.67}                                                       & \multicolumn{1}{c|}{1446}                                                  & \multicolumn{1}{c|}{1.62}                                                  & \multicolumn{1}{c|}{2/3/3/3/5}                                               & \multicolumn{1}{c|}{52.25}                                                     \\ \hline
\multicolumn{1}{|c|}{ALU}        & \multicolumn{1}{c|}{3983}                                                          & \multicolumn{1}{c|}{3835}                                                           & \multicolumn{1}{c|}{3.73}                                                       & \multicolumn{1}{c|}{1765}                                                  & \multicolumn{1}{c|}{0.03}                                                  & \multicolumn{1}{c|}{4/2/-/-/-}                                               & \multicolumn{1}{c|}{17.32}                                                     \\ \hline
\multicolumn{1}{|c|}{DCache}     & \multicolumn{1}{c|}{2175977}                                                       & \multicolumn{1}{c|}{2098008}                                                        & \multicolumn{1}{c|}{3.58}                                                       & \multicolumn{1}{c|}{2225}                                                  & \multicolumn{1}{c|}{7.8}                                                   & \multicolumn{1}{c|}{6/2/3/3/4}                                               & \multicolumn{1}{c|}{20.19}                                                     \\ \hline
\multicolumn{1}{|c|}{FPU}        & \multicolumn{1}{c|}{105770}                                                        & \multicolumn{1}{c|}{102140}                                                         & \multicolumn{1}{c|}{3.43}                                                       & \multicolumn{1}{c|}{1004}                                                  & \multicolumn{1}{c|}{0.51}                                                  & \multicolumn{1}{c|}{3/4/3/3/3}                                               & \multicolumn{1}{c|}{18.75}                                                     \\ \hline
\multicolumn{1}{|c|}{RegFile}    & \multicolumn{1}{c|}{110269}                                                        & \multicolumn{1}{c|}{106788}                                                         & \multicolumn{1}{c|}{3.16}                                                       & \multicolumn{1}{c|}{1384}                                                  & \multicolumn{1}{c|}{0.47}                                                  & \multicolumn{1}{c|}{5/3/2/3/3}                                               & \multicolumn{1}{c|}{24.27}                                                     \\ \hline
\multicolumn{1}{|c|}{FTQueue}    & \multicolumn{1}{c|}{191581}                                                        & \multicolumn{1}{c|}{185752}                                                         & \multicolumn{1}{c|}{3.04}                                                       & \multicolumn{1}{c|}{823}                                                   & \multicolumn{1}{c|}{2.51}                                                  & \multicolumn{1}{c|}{2/3/3/3/3}                                               & \multicolumn{1}{c|}{29.83}                                                     \\ \hline
\multicolumn{1}{|c|}{Rob}        & \multicolumn{1}{c|}{84443}                                                         & \multicolumn{1}{c|}{82188}                                                          & \multicolumn{1}{c|}{2.67}                                                       & \multicolumn{1}{c|}{816}                                                   & \multicolumn{1}{c|}{0.3}                                                   & \multicolumn{1}{c|}{2/3/3/3/3}                                               & \multicolumn{1}{c|}{31.32}                                                     \\ \hline
\multicolumn{1}{|c|}{BrPred}     & \multicolumn{1}{c|}{2126189}                                                       & \multicolumn{1}{c|}{2070836}                                                        & \multicolumn{1}{c|}{2.6}                                                        & \multicolumn{1}{c|}{1085}                                                  & \multicolumn{1}{c|}{6.69}                                                  & \multicolumn{1}{c|}{4/2/3/2/2}                                               & \multicolumn{1}{c|}{20.29}                                                     \\ \hline
\multicolumn{1}{|c|}{Rename}     & \multicolumn{1}{c|}{89945}                                                         & \multicolumn{1}{c|}{89297}                                                          & \multicolumn{1}{c|}{0.72}                                                       & \multicolumn{1}{c|}{325}                                                   & \multicolumn{1}{c|}{0.6}                                                   & \multicolumn{1}{c|}{2/3/3/-/-}                                               & \multicolumn{1}{c|}{4.43}                                                      \\ \hline
\multicolumn{1}{|c|}{Average}    & \multicolumn{2}{c|}{}                                                                                                                                                    & \multicolumn{1}{c|}{3.07}                                                       & \multicolumn{1}{c|}{1208.11}                                               & \multicolumn{1}{c|}{2.28}                                                  & \multicolumn{1}{c|}{}                                                        & \multicolumn{1}{c|}{24.29}                                                     \\ \hline
\multicolumn{8}{c}{Gemmini~\cite{gemmini}}                                                                                                                                                                                                                                                                                                                                                                                                                                                                                                                                                                                             \\ \hline
\multicolumn{1}{|c|}{Transposer} & \multicolumn{1}{c|}{2024}                                                          & \multicolumn{1}{c|}{1807}                                                           & \multicolumn{1}{c|}{10.72}                                                      & \multicolumn{1}{c|}{228}                                                   & \multicolumn{1}{c|}{0.43}                                                  & \multicolumn{1}{c|}{3/2/-/-/-}                                               & \multicolumn{1}{c|}{75.18}                                                     \\ \hline
\multicolumn{1}{|c|}{LoopConv}   & \multicolumn{1}{c|}{291632}                                                        & \multicolumn{1}{c|}{277801}                                                         & \multicolumn{1}{c|}{4.74}                                                       & \multicolumn{1}{c|}{1220}                                                  & \multicolumn{1}{c|}{3.27}                                                  & \multicolumn{1}{c|}{4/3/3/3/3}                                               & \multicolumn{1}{c|}{23.45}                                                     \\ \hline
\multicolumn{1}{|c|}{LpMatmul}   & \multicolumn{1}{c|}{76819}                                                         & \multicolumn{1}{c|}{74058}                                                          & \multicolumn{1}{c|}{3.6}                                                        & \multicolumn{1}{c|}{572}                                                   & \multicolumn{1}{c|}{0.36}                                                  & \multicolumn{1}{c|}{2/3/3/3/3}                                               & \multicolumn{1}{c|}{20.92}                                                     \\ \hline
\multicolumn{1}{|c|}{Spad}       & \multicolumn{1}{c|}{24771641}                                                      & \multicolumn{1}{c|}{24010083}                                                       & \multicolumn{1}{c|}{3.07}                                                       & \multicolumn{1}{c|}{1940}                                                  & \multicolumn{1}{c|}{99.68}                                                 & \multicolumn{1}{c|}{4/2/5/3/2}                                               & \multicolumn{1}{c|}{22.62}                                                     \\ \hline
\multicolumn{1}{|c|}{Mesh}       & \multicolumn{1}{c|}{147649}                                                        & \multicolumn{1}{c|}{144038}                                                         & \multicolumn{1}{c|}{2.45}                                                       & \multicolumn{1}{c|}{994}                                                   & \multicolumn{1}{c|}{0.83}                                                  & \multicolumn{1}{c|}{3/3/3/3/3}                                               & \multicolumn{1}{c|}{15.53}                                                     \\ \hline
\multicolumn{1}{|c|}{Average}    & \multicolumn{2}{c|}{}                                                                                                                                                    & \multicolumn{1}{c|}{4.91}                                                       & \multicolumn{1}{c|}{991}                                                   & \multicolumn{1}{c|}{20.91}                                                 & \multicolumn{1}{c|}{}                                                        & \multicolumn{1}{c|}{31.54}                                                     \\ \hline
\end{tabular}
	\label{resultTab}
\end{table}
	
	\subsection{Effectiveness of Proposed Pattern Mining Techniques}
	The experimental results are presented in Table~\ref{resultTab}. The major submodules of BOOM~\cite{boom} and Gemmini~\cite{gemmini} are evaluated individually. The submodules included in the table cover more than 75\% logic vertices in the gate-level netlist of BOOM or Gemmini. According to the "Pattern Cov(\%)"  in Table~\ref{resultTab}, a ratio showing how many vertices in the gate-level netlist are covered by the selected regular patterns, the existence of patterns is noticeable for most of the benchmarks since the average pattern coverage of the 31 benchmarks is 25.67\%.
	
	\subsubsection{Area Benefit} The resultant data of area are collected from the post-technology mapping reports before placement and routing. The area reduction ratio indicates how much area is reduced for a corresponding benchmark when pattern subgraphs in its gate-level netlist are replaced by the selected custom standard cells generated by AutoCellLibX. For benchmarks in EPFL benchmark suite~\cite{epfl}, submodules of BOOM~\cite{boom}, and submodules of Gemmini~\cite{gemmini}, their average area reduction ratios are 5.11\%,  3.07\%, and 4.91\% respectively. The average area reduction ratio for all the benchmarks in the table is 4.49\%. AutoLibCellX achieves significant area reduction compared to the FreePDK45 library while overlap constraint, pattern selection and area reduction are not considered in the previous works~\cite{AFSEM,FSMVLSI} of pattern mining for VLSI design.
	
	The optimization of area usage is noticeable but the customization benefit varies among benchmarks. Here, we analyze the related factors as follows:
	
	\begin{itemize}
		\item Design Hierarchy Depth: Benchmarks in EPFL benchmark suite~\cite{epfl} can get better results because they are designed specifically for a local circuit function so the patterns can have a higher coverage. In contrast, if AutoCellLibX consumes the entire gate-level netlist of BOOM or Gemmini without partitioning, each of the pattern groups will have a coverage lower than 10\%, and the resultant area reduction ratio will be lower than 1.5\% since some patterns are not shared by different submodules. Even submodules of these designs have multiple hierarchical levels. This indicates that proper partitioning of the input design can help to find out suitable custom standard cells. 
		\item Pattern Coverage and Pattern Size: Benchmarks like "ctrl", "i2c", and "Rename" benefits little from standard cell customization, because their netlists mainly consist of random logic so the pattern coverage is low. From another perspective, benchmarks like "mem\_ctrl" reach a lower area reduction ratio because the sizes of their high-frequent patterns are too small.
		
	\end{itemize}

	Compared to conventional frequent subgraph mining, for standard cell customization, AutoCellLibX emphasizes the overlaps between pattern subgraphs. For example, 3 of the 5 selected patterns for benchmark "DCache" overlap with each other on 49107 vertices in the gate-level netlist. Direct frequency-driven pattern mining, ignoring the overlaps, will lead to false frequency results while AutoCellLibX analyzes each vertex during neighbor encoding and pattern growth, avoiding overlaps and optimizing area usage.
	
	\subsubsection{Algorithm Efficiency} Considering neighbor encoding and pattern growth, the time complexity of pattern mining is reduced to $O(M(|E|+|V|))$, where $M$ is the maximum number of pattern mining iterations. According to the analysis of the termination criteria of pattern mining illustrated in Section~\ref{TerminationCriteriaofPatternMining}, for most scenarios, $M$ is less than 10. Therefore, our FSM algorithm is highly efficient, as shown in the "FSM time" in Table~\ref{resultTab}, and the time complexity of our FSM algorithm is much lower than the one of the gSpan algorithm~\cite{gspan} utilized in previous works~\cite{AFSEM,FSMVLSI}, which is proportional to the number of subgraphs in the gate-level netlist. However, currently, the timing-consuming layout synthesis, especially the simulated-annealing transistor placer and MILP compactor, takes up more than 95\% of the runtime of AutoCellLibX for the benchmarks, although AutoCellLibX has already pruned most of the minor patterns to avoid unnecessary layout generation and reduce the runtime. The average runtime is 1152 seconds for all the benchmarks while the longest runtime is 3869 seconds for benchmark "voter".
	
	\begin{figure}[t]
		\centering
		\includegraphics[width=0.8\linewidth]{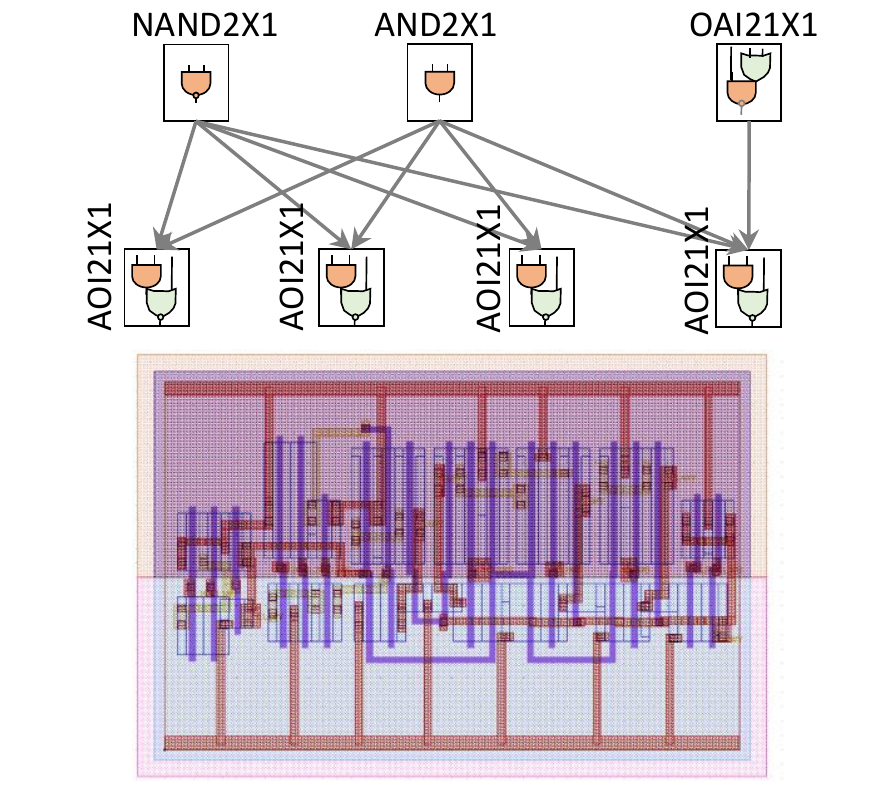}
		\caption{A large pattern subgraph with 7 standard cells for benchmark "router" and corresponding custom standard cell layout} 
		\label{largePattern}
	\end{figure}

	\begin{figure}[t]
		\centering
		\includegraphics[width=0.9\linewidth]{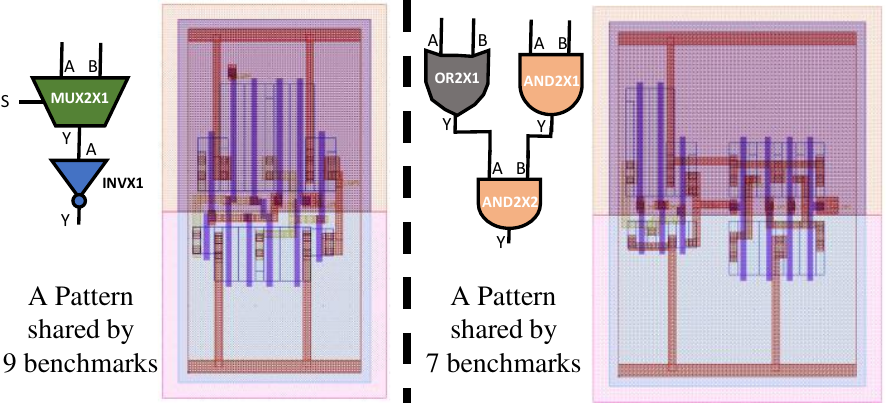}
		\caption{Some patterns shared across benchmarks in Table~\ref{resultTab}} 
		\label{sharedPatterns}
	\end{figure}

	\subsubsection{Characteristics of Custom Standard Cells}
	The "Pattern Size" column of Table~\ref{resultTab} shows the sizes of the selected pattern subgraphs for corresponding benchmarks. The largest pattern is the one with 7 vertices for benchmark "arbiter", as shown in Fig.~\ref{largePattern}, which it is hard for template-based algorithms~\cite{datapath1,datapath2,datapath3,datapath4,array1} to detect. Although we set $S_p$ to 10, the pattern mining procedures for all the benchmarks do not hit this bound which is predicted by Section~\ref{TerminationCriteriaofPatternMining}. Some benchmarks have less than 5 (i.e., $N_p$) custom standard cells since some candidates have been pruned due to low coverage during pattern growth. The patterns of benchmarks are various but some patterns are shared by some of the benchmarks, as shown in Fig.~\ref{sharedPatterns}, which shows the potential of reusing custom standard cells for different designs, improving productivity of VLSI design flow.
	
	\subsection{Existing Limitations and Future Works}
	
	While AutoCellLibX shows the power of efficient pattern mining for standard cell customization, it has several limitations, which are listed as follows:
	
	\begin{itemize}
		\item Currently, AutoCellLibX cannot the finish complete characterization of standard cells, e.g., the extraction of RC parasitics and timing information. While the results of area reduction are promising, the reward function $R(C,G)$ in Section~\ref{Preliminaries} can be replaced by a function of other factors like  delay and  power.
		\item Standard cell layout synthesis is the bottleneck of AutoCellLibX runtime. 
		 Moreover, the open-source layout generator ASTRAN~\cite{ASTRAN} does not support some of the design rules of high-end sub-10nm technology nodes currently. 
		Furthermore, factors considered in manual designs like  pin access, and via number reduction, could be included for practical application.
	\end{itemize}
	
	Since it is our initial attempt to explore the potentials of design-aware pattern mining of VLSI netlist and standard cell customization, these limitations are out of the scope of this proposed work, which presents an interesting and concrete approach to customizing standard cells, and we will consider more factors in our future works.
	
	 In the next development stage, AutoCellLibX will include the optimization of routability and timing and layout synthesis framework, to realize the practicality of AutoCellLib for sub-10nm process technology. Currently, we are adapting our algorithms to the technology design rules in ASAP7~\cite{ASAP7} and integrating the existing solutions from ~\cite{iTPlace,li2020mcell,DTCO,SPR,NCTUcell,7nm2019,Bonncell,PROBE2} into AutoCellLibX. Estimation model of timing, routability and DRC will be included in further evaluation. Any criticism and suggestions will be appreciated!

	\section{Conclusion}
	
	In this paper, we presents an automatic standard-cell library extension framework, AutoCellLibX. According to the post-technology mapping gate-level netlist of the design and the initial standard cell library, AutoCellLibX can find a set of standard cell cluster pattern candidates that occur frequently based on our proposed high-efficiency sub-graph mining algorithm for gate-level netlist. Meanwhile, to maximize the area benefit of standard cell customization for the given gate-level netlist, AutoCellLibX includes our proposed pattern combination algorithm which can iteratively find a set of gate-level patterns from numerous candidates as the extension part of the given initial library. AutoCellLibX closes the optimization loop between the analysis of gate-level netlist and standard cell library customization  for VLSI design productivity. The experiments with FreePDK45 library and benchmarks from various domains show that AutoCellLibX can generate the library extension with up to 5 custom standard cells in 1184 seconds on average for the benchmarks and the resultant extension of the standard cell library can save design area by 4.49\% averagely.

	\bibliographystyle{IEEEtran}
	\bibliography{sample-bibliography}

\end{document}